\documentclass[11pt,a4paper,oneside]{article}

\begin{document}

\title{Decay of light baryons by soft photon emission}

\author{T.\thinspace Jacobsen\\
Department of Physics, University of Oslo\\
PB 1048 Blindern\\
N-0316 Oslo, Norway}
\date{June 9, 2008}
\maketitle

\begin{abstract}
A possible reason for the emission of soft photons in high energy pp-collisions
is discussed.
\end{abstract}

\bigskip
\begin{flushleft}
\end{flushleft}
\bigskip
\bigskip

\noindent
Hadrons, i.e. baryons and mesons, have been extensively studied during
several decades. Because of their spacial extention, baryons are not point
particles but complex systems of some constituents taken to be three spin
1/2 quarks. Mesons are taken to be quark-antiquark systems with integer
spin.  According to the quark model, new baryons are formed if some quarks
of a nucleon are replaced by quarks with new ``flavours''.

The $\Lambda$(uds) at 1115 MeV is the lightest baryon which decays to 
a pion and a proton (uud) or a neutron (udd). If baryons too light for 
such decay exist, they must decay by emission of gammas. Production of 
Bremsstrahlung-like gammas not due to Bremsstrahlung has been seen in high 
energy proton-proton collisions \cite{bf}. Is it due to decay of low mass 
baryons? We have noticed that the ratio
\begin{eqnarray}
R(n) = m(n)/m(\mbox{nucleon}) 
\end{eqnarray}
between the observed mass $m(n)$ of some baryons and the mass $m(\mbox{nucleon})$
of the nucleon \cite{dpg} is fairly well reproduced by the formula
\begin{eqnarray}
m(n)/m(\mbox{nucleon}) = (1 + 1/n)^2
\end{eqnarray}
where $n \ge 1$ is an integer \cite{tj}, as if flavour and mass are not related, 
in disagreement with the standard model. These  hits, possibly 
accidental, suggest that these baryons are excited states of the nucleons since 
their masses are related to the mass of a nucleon by the same formula. If so, 
these baryons are excited states of the system of basic constituents of the 
nucleons.

For large and integer $n$, i.e. for $m(n)$ below threshhold for decay by pion 
emission, (2) gives 
\begin{eqnarray}
m(n) - m(\mbox{nucleon}) \approx  2 m(\mbox{nucleon})/n.
\end{eqnarray}

For increasing n and decreasing mass $m(n)$ the energy released by gamma-decay 
of $m(n)$ decreases while the density of states which decay increases, analogous 
to Bremsstrahlung. 

Vice versa, the observed production of Bremsstrahlung-like gammas not due to 
Bremsstrahlung seen in the above mentioned experiments \cite{bf} suggests that 
nucleons may be excited to states below threshhold for decay by emission of 
a pion. 

\bigskip
\begin{center}{\Large {\bf Acknowledgement}}\end{center}
\noindent
Discussions with K.M.\thinspace Danielsen, University of Oslo, Norway, are 
greatefully acknowledged.

\end{document}